\documentclass[aps,prb,twocolumn,superscriptaddress]{revtex4}
\usepackage{graphicx}
\usepackage{longtable}
\usepackage{array}
\usepackage{soul}
\usepackage{braket}
\usepackage{hyperref}
\usepackage{subfiles}
\usepackage{physics,bm}
\usepackage{xcolor}
\usepackage{float}
\newcolumntype{L}{>{\centering\arraybackslash}m{2cm}}
\newcolumntype{R}{>{\centering\arraybackslash}m{1.5cm}}
\newcolumntype{K}{>{\centering\arraybackslash}m{1.3cm}}

\begin{document}


\title{Nanometric modulations of the magnetic structure of the element Nd\footnote{This manuscript has been authored by UT-Battelle, LLC under Contract No. DE-AC05-00OR22725 with the U.S. Department of Energy.  The United States Government retains and the publisher, by accepting the article for publication, acknowledges that the United States Government retains a non-exclusive, paid-up, irrevocable, world-wide license to publish or reproduce the published form of this manuscript, or allow others to do so, for United States Government purposes.  The Department of Energy will provide public access to these results of federally sponsored research in accordance with the DOE Public Access Plan (http://energy.gov/downloads/doe-public-access-plan).}}

\author{H. Suriya Arachchige}
\email{ssuriyaa@vols.utk.edu}
\affiliation{Department of Physics \& Astronomy, University of Tennessee, Knoxville, TN 37996, USA}
\affiliation{Materials Science \& Technology Division, Oak Ridge National Laboratory, Oak Ridge, TN 37831, USA}

\author{L. M. DeBeer-Schmitt}
\affiliation{Neutron Scattering Division, Oak Ridge National Laboratory, Oak Ridge, TN 37831, USA}

\author{L. L. Kish}
\affiliation{Department of Physics, University of Illinois: Urbana Champaign, Urbana, IL 61801, USA}

\author{Binod K. Rai}
\affiliation{Materials Science \& Technology Division, Oak Ridge National Laboratory, Oak Ridge, TN 37831, USA}
\affiliation{Savannah River National Laboratory, Aiken, SC 29831, USA}
 
\author{A. F. May}
\affiliation{Materials Science \& Technology Division, Oak Ridge National Laboratory, Oak Ridge, TN 37831, USA}

\author{D. S. Parker}
\affiliation{Materials Science \& Technology Division, Oak Ridge National Laboratory, Oak Ridge, TN 37831, USA}

\author{G. Pokharel}

\affiliation{Department of Physics \& Astronomy, University of Tennessee, Knoxville, TN 37996, USA}
\affiliation{Materials Science \& Technology Division, Oak Ridge National Laboratory, Oak Ridge, TN 37831, USA}

\author{Wei Tian}
\affiliation{Neutron Scattering Division, Oak Ridge National Laboratory, Oak Ridge, TN 37831, USA}

\author{D. G. Mandrus}

\affiliation{Department of Physics \& Astronomy, University of Tennessee, Knoxville, TN 37996, USA}
\affiliation{Materials Science \& Technology Division, Oak Ridge National Laboratory, Oak Ridge, TN 37831, USA}
\affiliation{Department of Material Science \& Engineering, University of Tennessee, Knoxville, TN 37996, USA}

\author{M. Bleuel}
\affiliation{NIST Center for Neutron Research, National Institute of Standards and Technology, Gaithersburg, MD, 20899, USA}
\affiliation{Department of Materials Science and Engineering, University of Maryland, College Park, MD, 20742, USA}

\author{Z. Islam}
\affiliation{Advanced Photon Source, Argonne National Laboratory, Lemont, IL 60439, USA.}

\author{G. Fabbris}
\affiliation{Advanced Photon Source, Argonne National Laboratory, Lemont, IL 60439, USA.}

\author{H. X. Li}
\affiliation{Materials Science \& Technology Division, Oak Ridge National Laboratory, Oak Ridge, TN 37831, USA}

\author{S. Gao}
\affiliation{Materials Science \& Technology Division, Oak Ridge National Laboratory, Oak Ridge, TN 37831, USA}

\author{H. Miao}
\affiliation{Materials Science \& Technology Division, Oak Ridge National Laboratory, Oak Ridge, TN 37831, USA}

\author{S. M. Thomas}
\affiliation{MPA-Q, Los Alamos National Laboratory, Los Alamos, NM 87545, USA}

\author{P. F. S. Rosa}
\affiliation{MPA-Q, Los Alamos National Laboratory, Los Alamos, NM 87545, USA}

\author{J. D. Thompson}
\affiliation{MPA-Q, Los Alamos National Laboratory, Los Alamos, NM 87545, USA}

\author {Shi-Zeng Lin}
\affiliation{Theoretical Division, Los Alamos National Laboratory, Los Alamos, NM 87545, USA}

\author{A. D. Christianson}
\email{christiansad@ornl.gov}
\affiliation{Materials Science \& Technology Division, Oak Ridge National Laboratory, Oak Ridge, TN 37831, USA}

\date{\today}

\begin{abstract}
The rare earth neodymium arguably exhibits the most complex magnetic ordering and series of magnetic phase transitions of the elements. Here we report the results of small-angle neutron scattering (SANS) measurements as a function of temperature and applied magnetic field to study magnetic correlations on nanometer length scales in Nd.  The SANS measurements reveal the presence of previously unreported modulation vectors characterizing the ordered spin configuration which exhibit changes in magnitude and direction that are phase dependent.  Between 5.9 and 7.6 K the additional modulation vector has a magnitude $Q$ = 0.12 \AA$^{-1}$ and is primarily due to order of the Nd layers which contain a center of inversion.  In this region of the phase diagram, the SANS measurements also identify a phase boundary at $\approx$1 T. An important feature of these modulation vectors is that they indicate the presence of nanometer length scale spin textures which are likely stabilized by frustrated Ruderman-Kittel-Kasuya-Yosida (RKKY) interactions rather than a Dzyaloshinskii-Moriya (DM) exchange interaction. 
\end{abstract}

\maketitle

\section{Introduction}

Frustrated spin-spin interactions are the key to realizing a vast array of novel physical properties including spin liquids\cite{savary_2016,broholm_2020}, deconfined quasiparticles such as magnetic monopoles\cite{castelnovo_2008}, and novel types of magnetic order\cite{starykh_2015}. Of interest here is how such interactions generate spin textures far larger than a single chemical unit cell.  The most prominent example of a spin texture is a skyrmion\cite{back_2020}, a solitonic spin whirl, which has been observed to vary in size from 1.9 nm\cite{khanh_2020} to 1 $\mu$m\cite{nagaosa_2013}.  A defining feature of these structures are the topological properties which not only serve as a source of phase stability but also give rise to emergent electrodynamics which show promise for device applications\cite{fert_2017}.

\begin{figure}[h]
\includegraphics[width= 0.99\columnwidth] {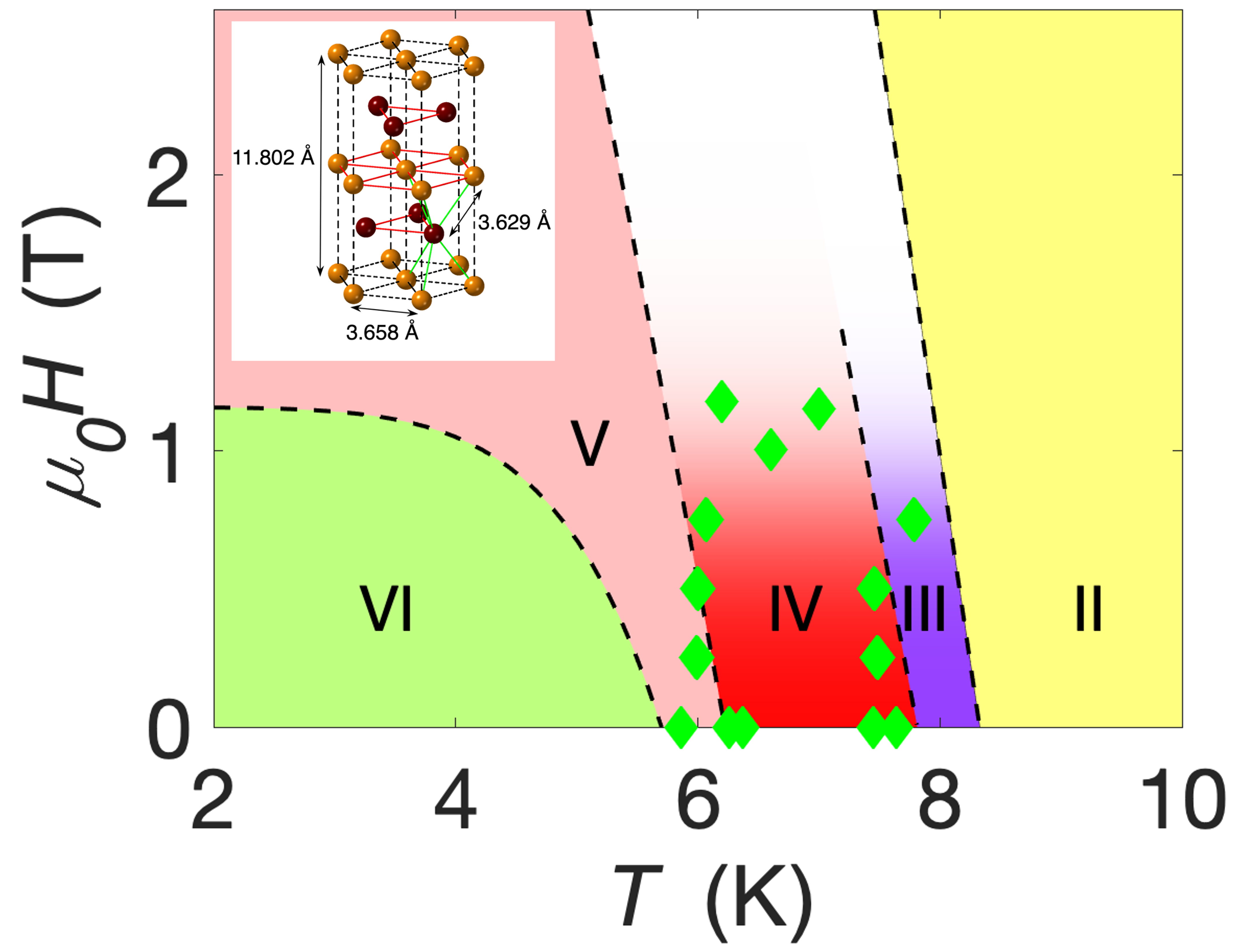}
\caption{\label{struct}  Field-temperature phase diagram ($\mu_0H$ $\parallel$ $c$). The dashed lines were determined from magnetostriction and thermal expansion measurements\cite{Zochowski1991-ag}. The green diamonds are determined from the SANS measurements reported here.  The individual phases are labeled by Roman numerals and are briefly described in the text. Phase I (not shown for clarity) exists between 19.9 and 19.1 K.   The inset shows the double hexagonal closed packed structure of Nd (space group $P6_3mmc$ (\#194)) with lattice parameters a = 3.658 \AA ~and c = 11.802 \AA.  The stacking sequence for this structure is ABAC. The two crystallographically distinct Nd sites are indicated by gold (cubic sites in the A layers) and maroon (hexagonal sites in the B and C layers).} 
\end{figure}


Understanding the underlying spin-spin interactions driving spin texture formation is an ongoing challenge for both theory and experiment.  However,  there have been recent theoretical advances in understanding the stabilization of skyrmions.  For instance, while canonical examples of skyrmion formation relied on a Dzyaloshinskii-Moriya (DM) exchange interaction in non-centrosymmetric materials,\cite{Muhlbauer2009-nz,Yu2010-gc,heinze_2011,nagaosa_2013,fert_2017,back_2020} concepts for\cite{Okubo2012-ah,leonov_2015,Lin2016-qp,Lin2018-qm,Hayami2016-qf,hayami_2017,PhysRevB.93.064430} and experimental realizations\cite{kurumaji_2019,hirschberger_2019,khanh_2020} of frustration stabilized skyrmions in centrosymmetric materials have recently been established.   The unique demonstration of skyrmion formation is challenging, in part due to complexities with domain effects and modelling of complex multi-\textbf{\textit{Q}} magnetism.  These two challenges come to head in elemental Nd, which is known to contain competing magnetic phases but also possesses the characteristics required to host frustration driven skyrmions.  To better refine theoretical models and assess the potential for Nd to host skyrmions, a more detailed characterization of the spin textures and how they evolve with applied field is required.  This work reports progress in such an effort, and in particular examines the scattering data at very low $Q$ and reveals additional spin modulations indicating spin textures with length scales of $\approx$5 nm.

Neodymium crystallizes in a double hexagonal closed pack (DHCP) structure (space group $P6_3/mmc$). The triangular nets formed by hexagonal planes of this crystal structure are stacked in an ABAC stacking sequence with two distinct Nd sites as shown in the inset of Fig. \ref{struct}.  The Nd sites in the A layers are typically referred to as cubic sites due to the approximate local face centered cubic symmetry, whereas the Nd sites in the B and C layers are typically referred to as hexagonal sites\cite{Moon1964-vg}. The symmetry elements of the cubic sites include a center of inversion whereas inversion symmetry is absent for the hexagonal sites. The magnetic properties of Nd have been generally discussed by considering the two Nd sites as largely decoupled\cite{Moon1964-vg,Lebech_1994}.

The magnetic properties of Nd are complex and still incompletely understood.  In the following, we provide a brief overview of the current understanding of the magnetic properties of Nd\footnote{Note there are small differences in the reported transition temperatures as well as thermal hysteresis, so for the sake of consistency we use the values reported in \cite{Zochowski1991-ag}.}.  Starting at 19.9 K, Nd undergoes a series of magnetic phase transitions as temperature is lowered.   Below 19.9 K, the hexagonal sites order in a  single-\textbf{\textit{Q}} antiferromagnetic structure\cite{Moon1964-vg,Koehler1965-jp,Lebech_1994} (phase I (not shown for clarity in Fig. \ref{struct})). The next transition occurs at 19.1 K to a two-\textbf{\textit{Q}} structure signified by the tangential splitting of the magnetic Bragg peaks\cite{moon1979does,lebech1979x} (phase II). The succeeding transition, which arises from ordering of the cubic sites (phase III), takes place at 8.2 K.   This phase is characterized by the appearance of magnetic Bragg peaks along the $\langle$1 0 0$\rangle$ directions with $|Q|$ = 0.3581 \AA$^{-1}$ $(0.18~\boldsymbol{a^*})$  \cite{Moon1964-vg,Lebech_1981}. The next transition occurs at 7.7 K (phase IV) with ambiguity as to the extent to which this phase persists under applied field; a schematic of the known modulation vectors is provided in Fig. \ref{schematic_peaks} in Appendix \ref{schematic_ps}.  At 6.3 K a transition to a proposed  four-\textbf{\textit{Q}} structure (phase V)  resulting from the increased coupling between the hexagonal and cubic sites\cite{Forgan1989-sl} is characterized by the vectors (0.106 0.00), (0.116 0.000), (0.181 0.013), (0.184  0.021) appearing around the (0 0 3) lattice point\cite{Forgan1989-sl}, where $Q_x$ is along $\langle$1 0 0$\rangle$ directions and $Q_y$ is perpendicular to $\langle$1 0 0$\rangle$ directions in the basal plane. In phases I-V, the moments lie within the basal plane.  Upon cooling below 6 K, the moments on the cubic site are no longer confined to the basal plane even in zero field (phase VI)\cite{gibbons_1992}.   In addition to the long-range ordered phases previously reported, the low-temperature phase has recently been proposed to exhibit properties of a self-induced spin glass \cite{kambereaay_2020}. Under applied magnetic fields, the phase diagram becomes considerably more complex and depends on the direction of the applied field \cite{Zochowski1991-ag}. Prior neutron scattering studies of the field-dependent magnetic properties have primarily focused on applied fields in the hexagonal planes\cite{Forgan1992-ss,mcewen_1990} and have utilized conventional neutron diffraction methods rather than small angle neutron scattering (SANS) measurements.

In this paper, we use SANS measurements to examine the ordered spin configuration in the element Nd. The key result presented here is the observation of additional magnetic Bragg peaks in the phases below 8 K indicating spin textures with large length scales up to 6.3 nm.  Between 5.9 and 7.6 K the new set of peaks is characterized by a modulation vector of 0.06 \textbf{a*} (0.12 \AA$^{-1}$) and hence a characteristic length scale of 5.2 nm. Below 5.9 K an additional set of peaks is observed, and at 2 K the modulation vectors are $(Q_x,Q_y) =(0.0839, 0.0601), (0.1093, 0.0928), (0.1442, 0.0424)$ and symmetry related positions\footnote{$Q_x$ is along $(1 0 0)$ and $Q_y$ is along $(\bar{1} 2 0)$}.  The results presented here show that the magnetic order in elemental Nd is more complex than previously appreciated and that Nd offers an important platform for exploring how large scale spin structures can emerge from interactions between localized moments and conduction electrons.

\begin{figure*}
\includegraphics[width= 2\columnwidth] {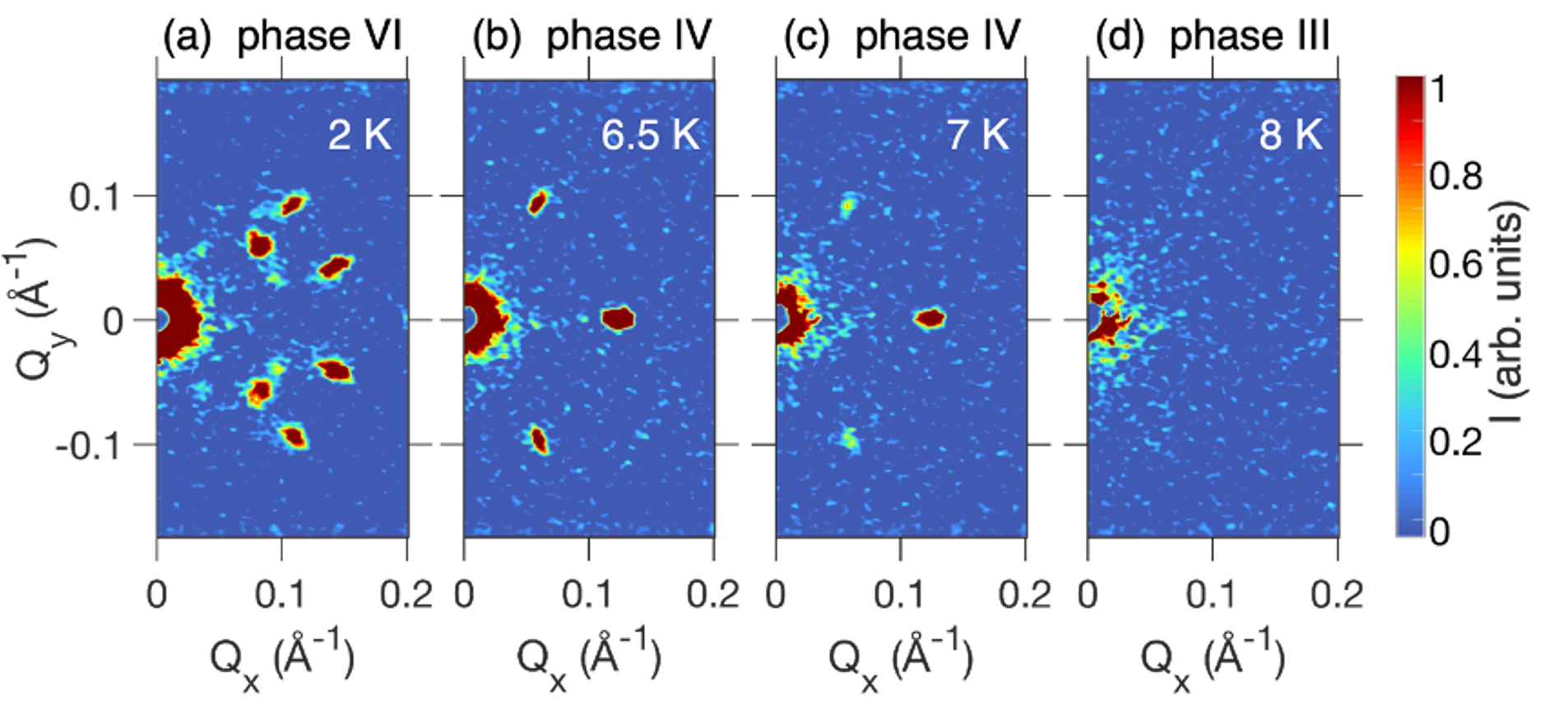}
\caption{
(a)-(d) display SANS data at 2, 6.5, 7, and 8 K respectively with $\mu_0H$ = 0. At 6.5 K (b) and 7 K (c) a set of peaks characterized by the modulation vector of 0.06 \textbf{a*} (0.12 \AA$^{-1}$) is observed. At 2 K, additional peaks are observed which can be indexed by modulation vectors of $(Q_x,Q_y)$ =(0.0839 \AA$^{-1}$, 0.0601 \AA$^{-1}$), (0.1093 \AA$^{-1}$, 0.0928 \AA$^{-1}$), (0.1442 \AA$^{-1}$, 0.0424 \AA$^{-1}$). The data were collected to optimize scattering for $Q_x \geq 0$. A full scattering pattern at 6.5 K is given in Fig. \ref{pretty_with_ferro}.  The roman numerals indicate the phases as labeled in Fig. \ref{struct}. $Q_x$ is along (1 0 0) and $Q_y$ is along (-1 2 0). }
\label{temp_dep}
\end{figure*}

\section{Experimental Details}

Small angle neutron scattering measurements were performed at the GP-SANS beamline at High Flux Isotope Reactor (HFIR) at ORNL and the NG-7 SANS at the NIST center for neutron research (NCNR). For the GP-SANS measurements, sample to detector distances of 3.5 and 3.0 m were used with neutron wavelengths of 4.75 and 4 \AA ~respectively. For these measurements a sample aperture diameter of 10 mm, a collimation aperture with a 40 mm diameter, and a 16.7 m aperture separation were used. To extract more precise correlation lengths at 6.5 K, high resolution scans were performed with a sample to detector distance of 3.56 m, a sample aperture with 8 mm diameter, and a collimation aperture with a 40 mm diameter.  The apertures were separated by a  distance of 12.93 m and wavelength of 4.75 \AA ~was used. A  wavelength spread $\Delta\lambda/\lambda=0.132$ was used in all of the GP-SANS measurements\cite{wignall201240,heller2018suite}. For the NG-7 SANS measurements, a sample to detector distance of 1 m, $\lambda$= 5 and 6 \AA, and $\Delta\lambda/\lambda=0.136$ were used. For these measurements, pieces cut from the same single crystal were used. The single crystal was grown via the Bridgeman technique by the material preparation center at Ames National Laboratory. The samples were mounted with the H0L plane horizontal. Magnetic fields were applied along the c-axis with a horizontal superconducting magnet. A schematic diagram of experimental setup is  provided in Fig. \ref{setup} of Appendix  \ref{app_setup}. Unless otherwise indicated, the measurement protocol for the SANS measurements reported here is zero field cooling the sample and collecting data on warming. After measurements involving applied magnetic fields, samples were warmed to 25 K (T$_\textit{N}$ = 19.9 K) before additional data was collected. For the analysis and data presented here, data collected at 25 K has been used as the background and subtracted from the data.

\begin{figure*}
\includegraphics[width= 1.8\columnwidth] {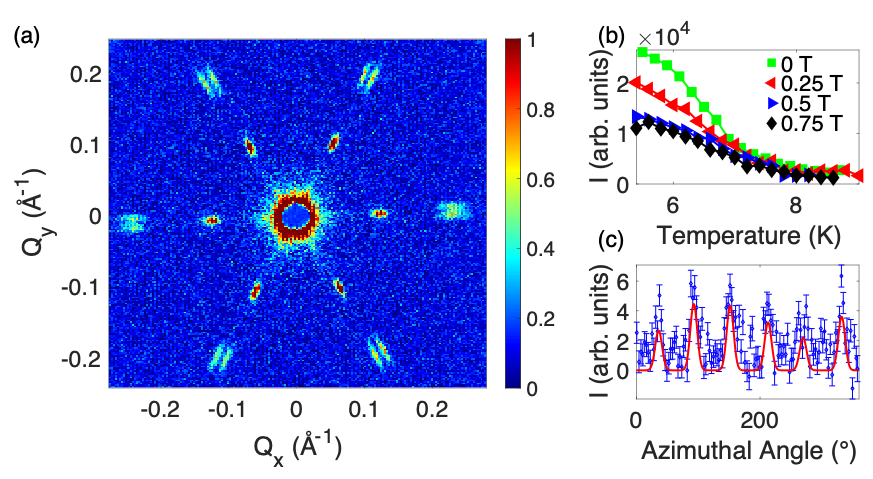}
\caption{\label{pretty_with_ferro}
(a) SANS pattern collected at 6.5 K. Peaks at $Q_{1}$ (inner set) and the set of peaks corresponding to the ordering of the hexagonal sites (outer set) are observed. (b) shows the temperature and field dependence of the scattering observed near $Q \approx 0$ by integrating the scattering bounded by concentric circles with radii 0.015 and 0.042 \AA$^{-1}$.  No significant changes in the background scattering were observed for the values of $T$ and $\mu_0H$ shown here.  Panel (c) shows the modulation of the radially integrated intensity $Q \approx 0$ as a function of the azimuthal angle (blue). The  azimuthal angle is plotted in the counterclockwise direction with the positive $Q_y$ axis as the starting point.  The red curve consists of six Gaussians constrained to the same width and a constant background.  The six peaks are located $\approx$ 60$^{\circ}$ apart, demonstrating that the low $Q$ scattering conforms the the hexagonal symmetry exhibited by the scattering at higher $Q$. }
\end{figure*}

\section{Results and Discussion}

\begin{figure*}
\includegraphics[width= 1.7\columnwidth] {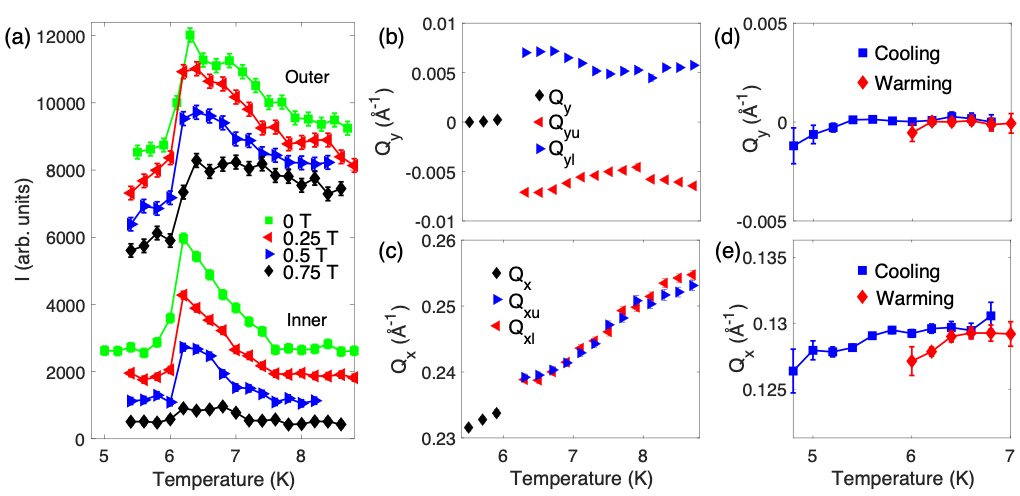}
\caption{\label{mesh}   
(a) Temperature dependence of the peaks observed at $Q_{1}$ (inner set) and at $Q \approx 0.23$ \AA$^{-1}$ (outer set) which are due to order of the hexagonal Nd sites.  The data were collected on warming.  The data are shifted by 700 units along the vertical axis for clarity with the 0.75 T data as the baseline at 700 units for \textit{both} the inner and outer sets of peaks.  Note that the intensity of the outer set of peaks remains significant throughout the temperature range shown here. (b) and (c) change in the $Q_y$ and $Q_x$ components of the outer peaks with increasing temperature respectively.  $Q_{x(y)}$ represents the position of the outer peak along $(100)$ ($(\bar{1}20)$). The subscripts \textit{u} and \textit{l} correspond to each peak after splitting. Panels (d) and (e) show the temperature dependence of the peak positions of the inner peaks. 
}
\end{figure*}

An overview of the SANS data as function of temperature at zero magnetic field is displayed in Fig. \ref{temp_dep} with additional details provided in Figs. \ref{pretty_with_ferro} and \ref{mesh}. As described further below, the changes in the scattering pattern as a function of temperature are consistent with the known phase diagram\cite{Zochowski1991-ag}. However, peaks corresponding to additional modulation vectors are observed in phases IV, V, VI.  Although the SANS data was collected on warming, we start our discussion of the data at higher temperature where the magnetic order and the corresponding SANS patterns are the least complex.

At 8 K the SANS data are essentially featureless (Fig. \ref{temp_dep}(d)).  This is consistent with the expectations of previous work where both the cubic sites and the hexagonal sites are already magnetically ordered, but with modulation vectors larger in magnitude than observable in the data presented in Fig. \ref{temp_dep}.  In particular, at 8 K, the smallest modulation vectors previously reported to characterize the spin configuration are 0.18 \textbf{a*} (0.358 \AA$^{-1}$) for the cubic sites and 0.12 \textbf{a*} (0.2387 \AA$^{-1}$) \cite{Lebech_1994} for the hexagonal sites and hence magnetic Bragg peaks in this temperature range lie outside of region of reciprocal space shown in Fig. \ref{temp_dep}(d). Figures \ref{temp_dep}(b) and (c) show that at 6.5 and 7 K (phase IV)  an additional set of peaks with a modulation vector of 0.0607(4) \textbf{a*} (0.1207(8) \AA$^{-1}$) are observed and are henceforth referred to as $Q_{1}$ peaks.  This scattering is then associated with a real space length scale of 5.2 nm, which will be discussed in additional detail below.

In phase IV, the ordered spin configuration has been previously described as a triple-\textbf{\textit{Q}} structure with two of the k-vectors ($Q\approx$ 0.12 \textbf{a*} (0.239 \AA$^{-1}$))  related to ordering of the hexagonal sites and the third ($Q$=0.18 \textbf{a*} (0.358 \AA$^{-1}$)) related to cubic site ordering\cite{Lebech_1994} (See also Fig. \ref{schematic_peaks} in Appendix \ref{schematic_peaks}). In Fig. \ref{pretty_with_ferro}(a), peaks corresponding to the hexagonal site order are observed as the outer set of peaks characterized by a modulation vector of 0.120(5) \textbf{a*} ($Q_x$=0.2391(5) and $Q_y$=0.071(1)) and peaks corresponding to the cubic order are outside of the $Q$-range probed by our SANS measurements.  The peaks due to hexagonal site order are weakly split in the direction transverse to $Q$ (the small $Q_y$ components given above) as expected based on studies in higher zones around the Bragg points (001) and (100) \cite{Lebech_1994}.  In contrast, the modulations characterizing the magnetic order of the cubic sites have not been reported to result in the peak splitting at 7 K\cite{Lebech_1994}. These observations can be utilized to infer the origin of the peaks at $Q_{1}$.   Since there is not any discernible splitting of the $Q_{1}$ peaks (See Appendix \ref{app_setup}, Fig. \ref{highres_roi} for a high resolution scan) this is strong evidence that the peaks do not originate from order solely due to the hexagonal sites, but rather stem from order on the cubic sites.  


Further support that the peaks at $Q_{1}$ are not a consequence of the hexagonal site order and hence originate from ordering of the cubic sites is provided by the temperature dependence of the scattering of the $Q_{1}$ peaks.  As shown in Fig. \ref{mesh}, monitoring the intensity of the $Q_{1}$ peaks as function of temperature (Fig. \ref{mesh}(a)) shows that the scattering intensity reaches its maximum at 6.2 K and there are no significant changes in the peak position as function of temperature (Fig. \ref{mesh}(d) and (e)).  This is in contrast to the behavior of the peaks which originate with the ordering of the hexagonal sites of the Nd structure where the peak position changes significantly in the same temperature range (Fig. \ref{mesh}(b) and (c)).  Additionally, the possibility of thermal hysteresis of the scattering at $Q_{1}$ was studied (See Appendix \ref{mag_thermal} Fig. \ref{hysteresis_neu}(a)).    The results are consistent with previous reports that the phase transitions in this temperature range exhibit thermal hysteresis\cite{Forgan_1979, Lebech_1994}.

An additional possible explanation of the peaks at $Q_{1}$ is that they are harmonic peaks arising from linear combinations of other fundamental wave vectors.  As has already been described, in phase IV there are distinct sets of modulation vectors characterizing the hexagonal site and cubic site order.  Hence, a potential explanation of the peaks at $Q_{1}$ is that magnetic order of the two sites is sufficiently coupled that the new peaks arise due to a linear combination of the vectors characterizing  hexagonal site order and/or cubic site order.  As already mentioned, the temperature dependence of the peaks at $Q_{1}$ and those due to hexagonal site order is distinct, such that the peaks due to cubic site order (outside of the SANS measurement window would also have to exhibit temperature dependence to compensate for this mismatch\cite{Lebech_1994}. This is contrary to the previously reported temperature dependence of the peaks due to cubic site order \cite{Lebech_1994}. Another possibility is that $Q_{1}$ = $\frac{1}{3}Q_{cubic}$.  Since the reported temperature dependence of the peaks at $Q_{cubic}$ appears to be similarly weak this is a strong indication that this relationship may be valid.   The simultaneous presence of two modulation vectors would then indicate that in Phase IV the cubic Nd sites exhibit multi-Q magnetic order--an important condition for the realization of a topologically non-trivial spin texture.  We further note that the peaks at $Q_{1}$ are absent above 7.6(1) K, whereas the order of the cubic site first occurs at 8.2 K\cite{Zochowski1991-ag,Lebech_1994}.  This indicates the modulation vector $Q_{1}$ stems from a change in the spin configuration of the cubic sites which introduces longer length scale modulations in phase IV than in phase III.

\begin{figure*}
\includegraphics[width= 2.1\columnwidth] {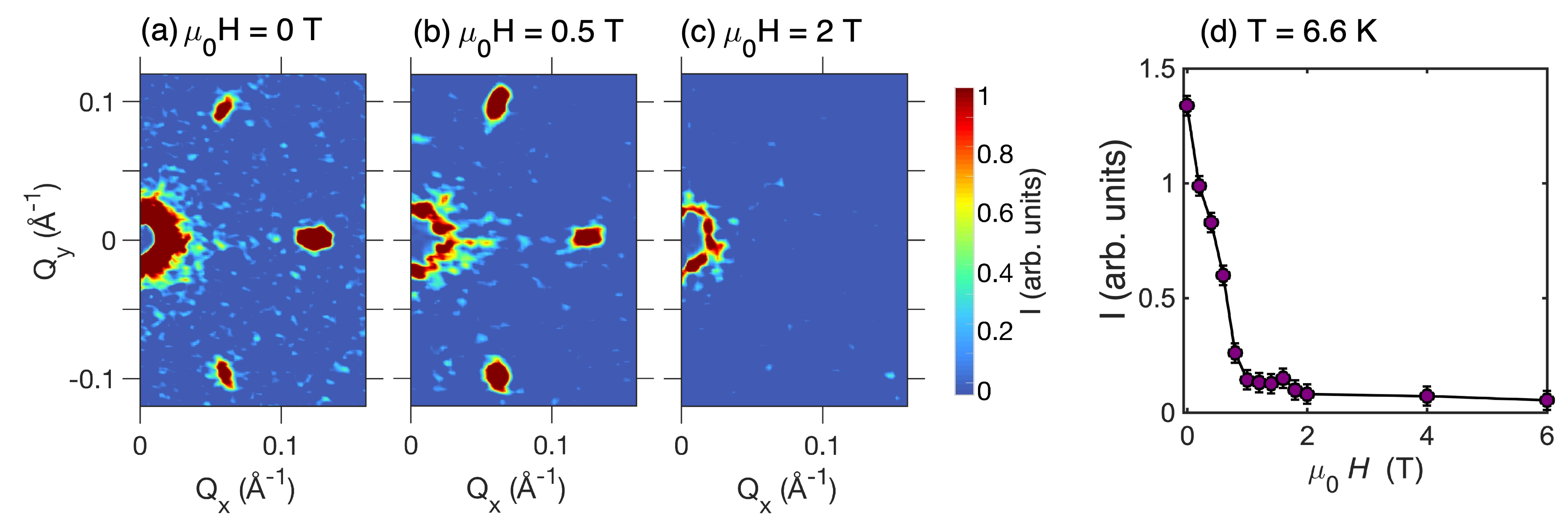}
\caption{
Field dependence of the scattering at $Q_{1}$. Panels (a)-(c) show the field-dependence ($\mu_0 H || c$) of the scattering at T = 6.5 K. The measurement protocol was zero field cooling the sample from 25 K to 2 K and   then warming under the specified field to 6.5 K.  Panel (d) shows the field dependence of the peaks at $Q_{1}$ at T = 6.6 K. For this measurement, the sample was zero field cooled to  6.6 K.}
\label{field_dep}   
\end{figure*}

Another interesting observation is that Fig. \ref{pretty_with_ferro}(b) shows that correlations around $Q = 0$ build at nearly the same time as the peaks at $Q_{1}$ appear.  The scattering around $Q = 0$ is likely an indicator of ferromagnetic correlations and this conclusion is further corroborated by a similar trend in the magnetization data (see Appendix \ref{mag_thermal}, Fig. \ref{dc_sus}).  The fact that the scattering is observable at finite $Q$ implies a short correlation length, unlike other magnetic scattering in the SANS data.  However, estimating a reliable correlation length is not possible since the majority of the scattering is obscured by the beam stop centered at $Q=0$.  Given this, this scattering may possibly be centered at finite Q rather than being centered precisely at $Q=0$.  The field dependence of the scattering around $Q = 0$ is shown in Fig. \ref{pretty_with_ferro}(b) indicating that the scattering decreases with applied field.  The technical limitations of detecting scattering around the direct beam prevent determining whether or not the scattering is moving toward $Q=0$ and out of the detection window of the experiments or if the field causes other more subtle changes to the spin configuration at the expense of the $Q=0$ scattering.  The scattering itself possesses the same hexagonal symmetry of the scattering as at larger $Q$s, as shown in Fig. \ref{pretty_with_ferro} (c) supporting the notion that the scattering is an intrinsic effect.

Although a more detailed understanding of the complex low temperature phases is beyond the scope of the present study, we now briefly discuss the scattering at low temperatures, with the primary objective of confirming that a distinct phase boundary exists between the phase characterized by the peaks at $Q_{1}$ (IV) and phases V and VI.  Figure \ref{temp_dep}(a) clearly shows that several new peaks are evident at 2 K but that scattering above background at $Q_{1}$ is absent. The peaks at 2 K are indexed by $(Q_x,Q_y)$ =(0.0839 \AA$^{-1}$, 0.0601 \AA$^{-1}$), (0.1093 \AA$^{-1}$, 0.0928 \AA$^{-1}$), (0.1442 \AA$^{-1}$, 0.0424 \AA$^{-1}$) and symmetry related positions as shown in Fig. \ref{temp_dep}(a) (Recall: $Q_x$ is along $(1 0 0)$ and $Q_y$ is along  $(\bar{1} 2 0)$). The temperature dependence of the intensity of the newly observed low angle peaks of phase VI is shown in Appendix \ref{ap_low_temp_scatt}, Fig. \ref{phase_vi_t_dep}. The low temperature peaks disappear just as the peaks at $Q_{1}$ begin to appear with rising temperature.  The origin of these additional peaks is uncertain but may be related to harmonics or intermodulation harmonics of the fundamental vectors characterizing the order as the magnetism on the cubic and hexagonal sites is fully coupled in phase V and VI \cite{mcewen_1990,gibbons_1992,Forgan1992-ss,Forgan1989-sl}. Additional work is underway to understand the low temperature phases in greater detail.

Having established that the peaks at $Q_{1}$ indicate a newly detected modulation vector characterizing phase IV and likely stem from the ordered spin configuration of the cubic sites, we now examine the field dependence. Figure \ref{field_dep}(a)-(c) shows the response to a magnetic field applied along the c-axis. Fig. \ref{field_dep}(d) shows the combined integrated intensity of three of the peaks at $Q_{1}$ as a function of applied field at 6.6 K. The scattering pattern disappears at fields over 1 T. This field dependence is order parameter-like, indicating that there is a previously unreported phase boundary at $1.0(1)$ T. Although, the phase above 1 T is apparently the same as phase III, this is not conclusively demonstrated with the present data.   The outer set of peaks remain in the data above 1 T, but additional measurements are required to determine the relationship between the phase above 1 T and phase III.

We now discuss two length scales associated with the peaks at $Q_{1}$: 1) the corresponding spatial extent or size of the spin texture and 2) the range over which the spin texture is correlated.  As noted the length scale of the spin textures characterized by the peaks at $Q_{1}$ is 5.2 nm ($\approx$14 times larger than the in-plane lattice constant).  The other key length scale is characterized by the correlation length, $\xi$.  In this case, correlation lengths corresponding to the three spatial dimensions can be extracted (see Appendix \ref{cor_sect} for additional details).  In a polar coordinate system these are:  radial, $\xi_{Q}$ = 228 \AA , azimuthal, $\xi_{\phi}$ = 425 \AA, and polar $\xi_{\omega}$ = 1923 \AA. For comparison, the correlation lengths extracted for the outer set of peaks is given along with the correlation lengths for the scattering at $Q_{1}$ in Table \ref{correlation_table_short}.   The two sets of correlation lengths are largely consistent, though the correlation lengths for the peaks at $Q_{1}$ are about a factor of 2 larger for the radial and azimuthal directions.   The relatively short radial correlation length--though still a factor of 4 larger than the characteristic length scale--is considerably shorter than the other correlation lengths. This implies a degree of disorder involved with the primary wave-vector in the modulation, such as may result from strong interactions with the lattice through mechanisms such as pinning.  Note that sample mosaic would more strongly affect the correlation lengths in the azimuthal and polar directions rather than the radial correlation length.

An additional observation concerning the interaction of the magnetic properties with the lattice comes from attempts to measure a topological Hall effect. To do this, we measured the Hall effect in mechanically thinned samples of Nd. Initially, we mechanically thinned a sample to 15~$\mu$m with aim to fabricate a Hall bar using focused-ion-beam milling. In this 15~$\mu$m sample, however, we observed that the magnetic transition near 1 T at 2 K (Phase VI) was extremely broadened and occurred over a field range of 0.8 T (data not shown here). We then measured a mechanically thinned 225~$\mu$m sample instead, which also showed a broadened transition occurring over a field range of 0.4 T. No measurable signal for the topological Hall effect was found in either sample. The appearance of the topological Hall effect requires non-coplanar spin textures and its magnitude depends on several factors such as the size of the spin texture and the coupling between the conduction electrons and spin textures. The lack of a topological Hall effect within our experimental resolution may then imply an extremely weak Hall signal due to a non-coplanar spin texture, the sample properties are sufficiently changed by the mechanical thinning process, or possibly that a more complex spin texture is present that wouldn't exhibit a topological Hall effect\cite{afskyrmion_2017}.   We also examined the field dependence of the (220) structural Bragg peak with synchrotron x-ray diffraction on a single crystal (see Fig. \ref{xray} of Appendix \ref{x_raya}).  These results demonstrate a significant response by the structural peak near the phase transition at 1 T in phase IV.


\begin{table}
\renewcommand{\arraystretch}{1.4}
\caption{\label{correlation_table_short} Correlation lengths extracted as described in Appendix \ref{cor_sect}. $\xi_{Q}$, $\xi_{\phi}$, and $\xi_{\omega}$ indicate the radial, azimuthal, and polar correlation lengths respectively.}
\setlength{\tabcolsep}{7pt}
\begin{tabular}{c|cccccccccc}
\hline \hline
& $\xi_{Q}($\AA$)$&$\xi_{\phi}($\AA$)$&$\xi_{\omega}($\AA$)$ \\ 
\hline
$Q_{1}$ & 228 & 425 & 1923\\  
\hline
Outer Peak & 80 &  265 &  2173\\
\hline \hline
\end{tabular}
\end{table}

Despite the challenges in observing a topological Hall effect, an interesting possible explanation of the SANS data is that the scattering arises from the presence of a topologically nontrivial spin texture such as a skyrmion lattice in Nd. As the scattering appears to originate from the order on the cubic sites which have inversion symmetry this suggests that the mechanism is frustration driven rather than the more commonly observed spin textures driven by a DM exchange interaction.  Furthermore, as the magnetic order is already multi-$Q$ in Phase IV, the additional vector $Q_{1}$ is also likely the result of a multi-$Q$ spin arrangement.    In a metal such as Nd, the regions of high spin susceptibility originate from the conduction electron mediated Ruderman-Kittel-Kasuya-Yosida (RKKY) interactions\cite{fleming_1968,fleming_1969} which determines the ordering wave vectors. The understanding of the RKKY interactions requires knowledge of the Fermi surface as well as the hybridization function between the localized spin and conduction electrons, which have not yet been determined.

To proceed, we will propose possible magnetic interactions based on the experimental observations. Such an approach will provide a theoretical description of the experiments and will constrain future microscopic theories. Because the RKKY interactions are highly non-local in real space, it is more convenient to write the magnetic interactions in the momentum space. For a spin-rotation invariant system, the Hamiltonian to the second order in spin $\mathbf{S}$ is $\mathcal{H}_2=\sum_Q J_2(\mathbf{Q})\mathbf{S}(\mathbf{Q})\cdot\mathbf{S}(-\mathbf{Q})$, subjected to the local constraint $|\mathbf{S}(r)|=S_0$ with $S_0$ the size of the magnetic moment. The ground state spin configuration of $\mathcal{H}_2$ is a magnetic spiral with an ordering wavevector $Q_s$ minimizing $J_2(\mathbf{Q})$. Multi-\textbf{\textit{Q}} magnetic order is not favored by $\mathcal{H}_2$ because of the violation of the constraint $|\mathbf{S}(r)|=S_0$. This violation generates $S(\mathbf{Q})$ at higher order harmonics, $Q=2Q_s, 3Q_s,\ ...$, which costs energy. Other magnetic interactions such as easy axis anisotropy \cite{leonov_2015,PhysRevB.93.064430,PhysRevLett.124.207201}, compass interactions \cite{PhysRevB.103.104408}, and four spin interactions \cite{doi:10.7566/JPSJ.85.103703,hayami_2017} were demonstrated to stabilize multi-\textbf{\textit{Q}} magnetic textures. The four spin interaction is $\mathcal{H}_4=\sum_{Q_i} J_4(\mathbf{Q}_i)[\mathbf{S}(\mathbf{Q}_1)\cdot\mathbf{S}(\mathbf{Q}_2)][\mathbf{S}(\mathbf{Q}_3)\cdot\mathbf{S}(\mathbf{Q}_4)]$, which arises at quartic order in the perturbation theory. The translational invariance imposes $\mathbf{Q}_1+\mathbf{Q}_2+\mathbf{Q}_3+\mathbf{Q}_4=0$. The appearance of the triangle lattice of spin texture (triple-\textbf{\textit{Q}} order) indicates a condensation of $\mathbf{S}(\mathbf{Q}_i)$ at three $\mathbf{Q}_i$s with $\mathbf{Q}_1+\mathbf{Q}_2+\mathbf{Q}_3=0$. The triple-\textbf{\textit{Q}} order is favored when the four spin interaction is attractive, i.e. $J_4(\mathbf{Q}_i)<0$, and in the presence of a uniform magnetization, $\mathbf{S}(\mathbf{Q}=0)\neq 0$. The existence of a uniform magnetization component in the triple-\textbf{\textit{Q}} magnetic texture at zero magnetic field in Nd (see Fig. \ref{mesh}(a)) suggests the importance of $\mathcal{H}_4$ for stabilizing the triple-\textbf{\textit{Q}} magnetic texture.  Despite the above considerations, we stress the work presented here does not provide definitive evidence for a topologically nontrivial spin texture such as a skyrmion.  However, there is clearly a need for further work to account for additional modulation vectors to explain the complex magnetism exhibited by the element Nd.

\section{Conclusions}

We have studied the magnetic order in the element Nd with SANS measurements.  These studies have revealed an additional set of modulation vectors. In particular,  we have demonstrated the appearance of a 0.06 \textbf{a*} (0.12 \AA$^{-1}$) modulation vector at zero magnetic field between 5.9 and 7.6 K which appears to be due to the spin configuration of the Nd atoms occupying the cubic sites of the DHCP structure.  By tracking the temperature and field behavior of the scattering at $Q_{1}$, we have determined an additional phase boundary in Nd at 1 T. An important feature of findings is the presence of nanometer length scale spin textures ranging from 4.2 to 6.3 nm which are likely stabilized by frustrated RKKY mediated spin-spin interactions.  Our results show that the already complex magnetic behavior of Nd is richer than previously appreciated and additional investigations are likely to yield further insights into the fascinating behavior of the element Nd.

\begin{acknowledgments}
We thank C. Batista for useful discussions. This work was supported by the U.S. Department of Energy, Office of Science, Basic Energy Sciences, Materials Sciences and Engineering Division. This research used resources at the High Flux Isotope Reactor, a Department of Energy (DOE) Office of Science User Facility operated by Oak Ridge National Laboratory (ORNL). H. S. A. was supported by the Gordon and Betty Moore Foundation's EPiQS Initiative through Grant GBMF9069 and the Shull Wollan Center Graduate Research Fellowship.  G. P. was supported by the Gordon and Betty Moore Foundation's EPiQS Initiative through Grant GBMF4416. The work at LANL was carried out under the auspices of the US DOE NNSA under Contract No.89233218CNA000001 through the LDRD Program.  This research used resources of the Advanced Photon Source, a U.S. Department of Energy (DOE) Office of Science user facility at Argonne National Laboratory and is based on research supported by the U.S. DOE Office of Science-Basic Energy Sciences, under Contract No. DE-AC02-06CH11357.
\end{acknowledgments}


\clearpage

\onecolumngrid
\appendix

\section{Additional experimental details for the SANS measurements\label{app_setup}}

\renewcommand{\thefigure}{A\arabic{figure}}
\renewcommand{\thetable}{A\Roman{table}}
\setcounter{figure}{0}
\setcounter{table}{0}

\begin{figure}[h]
\includegraphics[width= 0.5\columnwidth]{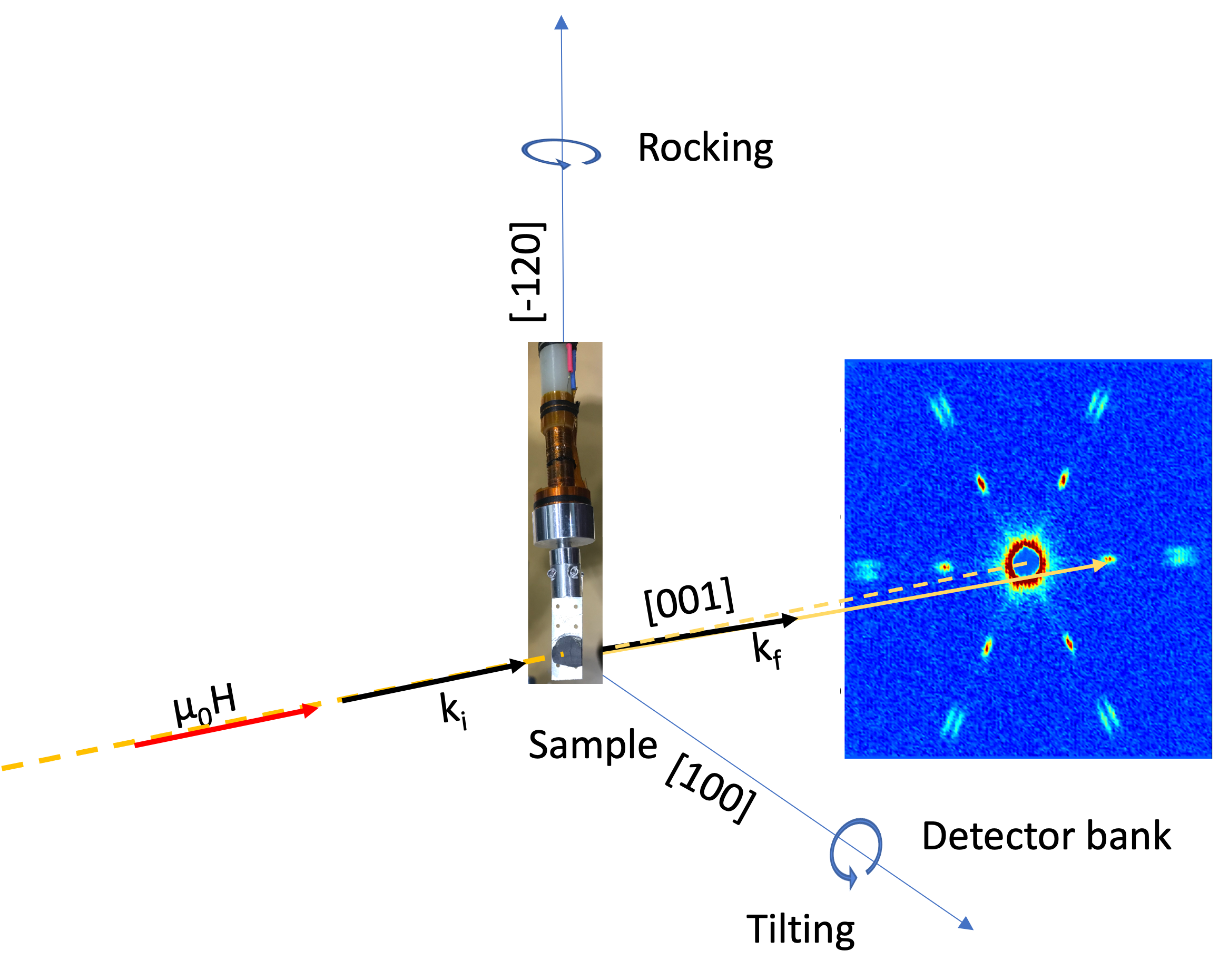}
\caption{Schematic diagram of the experimental set up for the SANS measurements. The magnetic field is applied along the [0 0 1] direction. The incoming neutron beam ($k_i$) is indicated by the yellow dashed line. For rocking curves the sample is rotated around the [$\bar{1}$ 2 0] direction, and for tilt scans around the [1 0 0] direction. The scattered neutrons ($k_f$) are then detected by a two dimensional detector bank.
\label{setup} }
\end{figure}

\begin{figure}[h]
\includegraphics[width= 0.5\columnwidth] {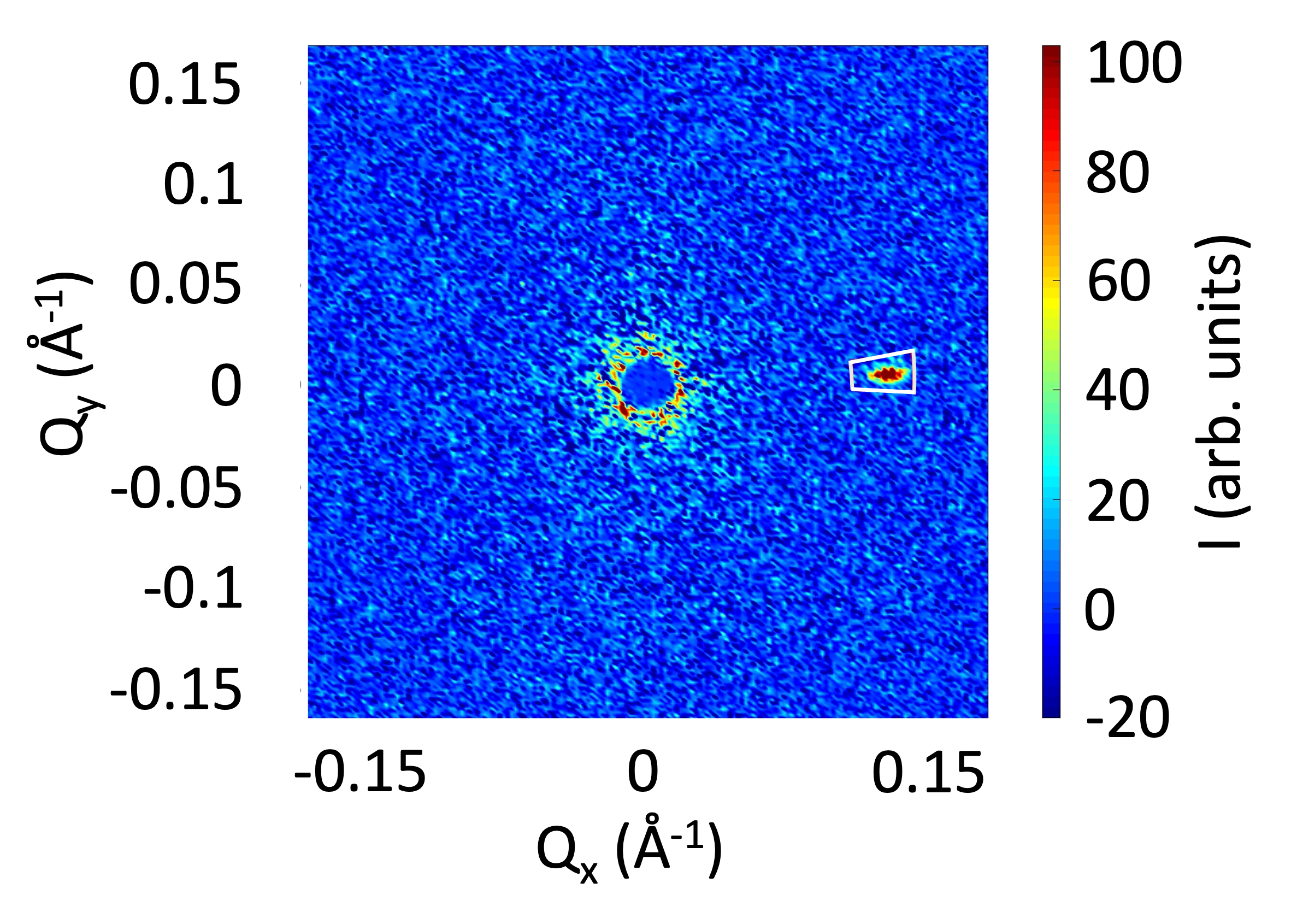}
\caption{\label{highres_roi}  High resolution SANS data. To extract precise correlation lengths, high resolution scans were performed on the right equatorial peak of the hexagonal scattering pattern ($Q_{1}$ modulations) appearing in Phase IV. The sector marked in white indicates the region of interest used to extract correlation lengths using the procedure described below. These high resolution scans do not evidence any peaks splitting.
}
\end{figure}

 As described in the main text, SANS measurements were carried out to probe the magnetic scattering with emphasis on Phase IV (Fig. \ref{struct}). A schematic of the experimental geometry is shown in Fig. \ref{setup}. In this experimental setup the field is applied along the c-axis of the sample and the sample is initially orientated such that the (h k 0) plane is perpendicular to incident neutron beam.  Rocking scans are then performed by rotating the sample about the vertical [1 -2 0] direction. For the SANS measurements, the sample was glued to an aluminum plate.  A high resolution scan used to provide an additional, more detailed, check of the correlation lengths of the peaks at $Q_{1}$ is shown in Fig. \ref{highres_roi}. The high resolution scans also provide further confirmation that there is no discernible peak splitting of the peaks at $Q_{1}$ such as observed for the peaks originating from order on the hexagonal Nd sites.

\section{Low temperature scattering}\label{ap_low_temp_scatt}

\renewcommand{\thefigure}{B\arabic{figure}}
\renewcommand{\thetable}{B\Roman{table}}
\setcounter{figure}{0}
\setcounter{table}{0}

\begin{figure}[h]
\includegraphics[width= 0.5\columnwidth] {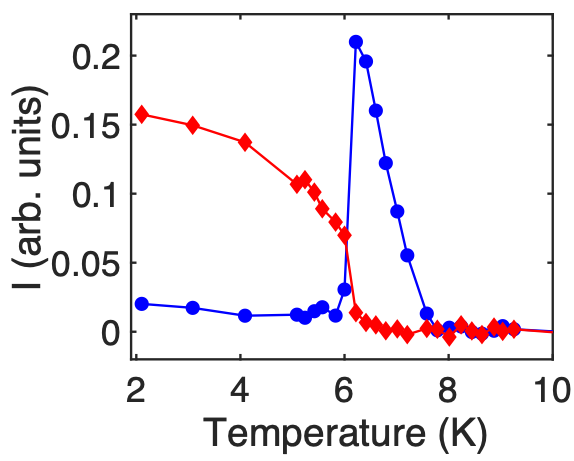}
\caption{\label{phase_vi_t_dep} Temperature dependence of the scattering at low temperature.  The red curve corresponds to peaks appearing in Phase VI as shown in Fig. \ref{temp_dep}(a) and the blue curve for the $Q_{1}$ modulations from Phase IV. The intensity of the red curve is the sum of the integrated intensities of the peaks appearing at 
$(Q_x,Q_y)$ = (0.084(1), 0.059(1)) \AA$^{-1}$;
(0.110(0), 0.093(9)) \AA$^{-1}$; (0.144(6), 0.041(8)) \AA$^{-1}$ as shown in Fig.\ref{temp_dep}(a). Near 6 K, coexistence of the modulation vectors is evident. Data for both curves were collected while warming.  
 }
\end{figure}

At 2 K (Phase VI) a complex scattering pattern is observed as shown in Fig. \ref{temp_dep}(a).  The complexity in the SANS pattern is in addition to the complexity observed previously with conventional neutron diffraction measurements which observed a complex scattering pattering at higher $Q$s\cite{Lebech_1994}. The complexity of the scattering  observed in Phase VI has been explained as a consequence of the magnetic interactions between the cubic and hexagonal orderings. This was reported to cause the $\approx$ 0.11 a$^*$ modulations corresponding to the ordering of the hexagonal sites to longitudinally split into two modulations,  and the $\approx$ 0.18 a$^*$ modulations corresponding to the ordering of the cubic sites to azimuthally split into four modulations\cite{Lebech_1994} around the (0 0 1) lattice point. Figure \ref{phase_vi_t_dep} shows the temperature dependence of the scattering pattern shown in Fig. \ref{temp_dep}(a) (characteristic of phase VI) at zero applied magnetic field compared to scattering of the $Q_{1}$s which are shown in \ref{temp_dep}(b) (characteristic of Phase IV). Phase V occurs between phases VI and IV, however there is no discernible feature of the present data set of specifically characteristic of phase V.   Near the phase boundary, there is a narrow temperature interval where scattering from features observed in phases VI and IV coexist, indicating the first order nature of this phase boundary as indicated in Ref. \cite{Lebech_1994}. This observation is also supported by the significant hysteresis observed at the lower temperature phase boundary of Phase IV as shown in Fig. \ref{hysteresis_neu}

\section{Correlation length extraction from the SANS measurements\label{cor_sect}}

\renewcommand{\thefigure}{C\arabic{figure}}
\renewcommand{\thetable}{C\Roman{table}}
\setcounter{figure}{0}
\setcounter{table}{0}

In magnetic neutron scattering, the shape of a Bragg peak is affected by a variety of factors both intrinsic to the sample and introduced by the instrument. Intrinsically, the shape of the scattering function is largely determined by the extent of magnetic correlations in the sample along each crystal direction. As a function of real-space distance $r$, two-point spin correlations are often assumed to follow an exponential decay behavior: 

\begin{equation}
\langle S(0)\cdot S(r) \rangle \propto e^{-r/\xi} 
\end{equation}
Here, the angle brackets denote thermal averaging in time and across the sample and the exponential decay is parametrized by a characteristic correlation length $\xi$. If the real-space periodicity leading to the reflection varies throughout the sample, such as due to disorder, the peak will broaden in the radial ($\hat{Q}$) direction. On the other hand, for crystals with a measurable mosaicity, the peak shape in the azimuthal and polar angular directions ($\hat{\phi}$,$\hat{\omega}$) will be convoluted with an angular probability distribution owing to the misalignment of different crystallites within the sample.

\begin{figure}
\includegraphics[width= 1\columnwidth] {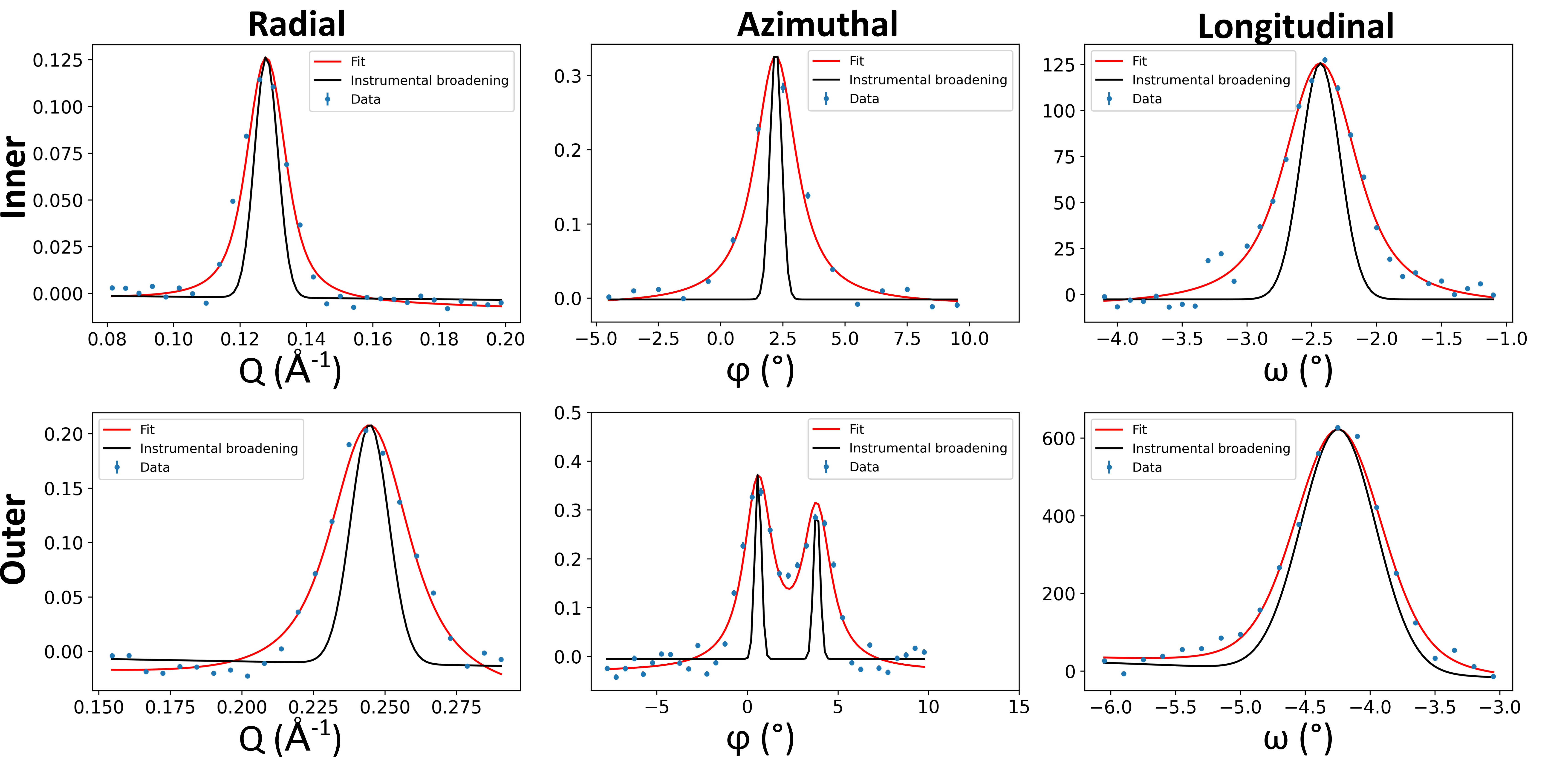}
\caption{\label{resolution_plots} Fits to peaks at $Q_{1}$ (top row) and peaks due to magnetic order of the hexagonal sites (bottom row) to determine the spin-spin correlation lengths. The black lines indicate the calculated instrumental resolution function as described in the text.  The red lines are the result of a fit of a Voigt function with the Gaussian component constrained to instrumental resolution.  The fitting parameters and correlation lengths are given in Table \ref{correlation_table}.
}
\end{figure}

Furthermore, the scattering function is always convoluted with the resolution function of the instrument. At GP-SANS the known resolution function at each $\vec{Q}$ is well-described by a three-dimensional Gaussian distribution\cite{cubitt_1992,mildner_1984}. The squared widths in each direction ($\sigma_{Q,\phi,\omega}$) are given by the equations: 

\begin{equation}
\sigma_Q^2= \left(\frac{k Q\Delta\lambda}{4\pi\sqrt{6}}\right)^2+\left(\frac{k\sqrt{r_1^2+r_2^2}}{l}\right)^2
\end{equation}

\begin{equation}
\sigma_\phi= \left(\frac{k\sqrt{r_1^2+r_2^2}}{l}\right)
\end{equation}
\begin{equation}
\sigma_Q^2= \left(\frac{Q^2\Delta\lambda}{4\pi\sqrt{6}}\right)^2+\left(\frac{Q\sqrt{r_1^2+r_2^2}}{l}\right)^2
\end{equation}

In our fits to experimental SANS data, we assume peak-widths dominated by exponential real-space correlations and by instrumental resolution. In this case, the peak-shape along each direction follows a Voigt profile ($f$), which is the convolution of the Lorentzian correlation and Gaussian resolution components: 

\begin{equation}
f(x;A,\mu,\sigma,\gamma)= A Real(\exp{(\frac{x-\mu+i\gamma}{\sigma\sqrt{2}})^2}\Gamma(-i(\frac{x-\mu+i\gamma}{\sigma\sqrt{2}})^2))
\end{equation}
Here $\Gamma$ is the error function,  $\sigma$ is equal to the standard deviation of the Gaussian (instrumental resolution) component in the Voigt function, and $\gamma$ gives the half-width of the Lorentzian component, which is inversely related to the correlation-length in our model: 
\begin{equation}
\xi=1/\gamma
\end{equation}

By fitting the peak profiles with a Voigt function and fixing the Gaussian width to the known value for the resolution function, we extracted magnetic correlation lengths for the equatorial peaks at $Q_{1}$ (inner peak) and peaks due to hexagonal site order (outer peak) in our SANS measurement.  The panels in Figure \ref{resolution_plots} show the variation of the integrated intensity in each of three directions in a spherical coordinate system (radial, azimuthal, and longitudinal/polar). The red curves show fits to Voigt profiles, with the fixed Gaussian resolution component illustrated by the black curve. Along polar and azimuthal directions, the data were fit in angular units but converted into reciprocal-space chord lengths for the correlation calculation. Table \ref{correlation_table} lists the obtained width parameters and estimated correlation lengths in each direction to the nearest Angstrom.

\begin{table*}
\renewcommand{\arraystretch}{1.4}
\caption{\label{correlation_table}   Fit parameters used in determining equatorial peak correlation lengths for peaks at $Q_{1}$ (Inner Peak) and peaks originating from magnetic order of the hexagonal Nd sites (Outer Peak). The subscripts $Q$, $\phi$, $\omega$ refer to fits along the corresponding polar directions. $d$ is the real-space distance associated with the peak-center and $\sigma$ and $\gamma$  are the Gaussian and Lorentzian width parameters within the Voigt function. As described in the text, $\sigma$ is fixed to the resolution value for extraction of the correlation length. $\xi$ is the real-space correlation length for each direction obtained from $\gamma$.}
\setlength{\tabcolsep}{7pt}
\begin{tabular}{c|cccccccccc}
\hline \hline
&$d$(\AA)& $\gamma_{Q}($\AA$^{-1})$&$\sigma_Q($\AA$^{-1})$& $\xi_{Q}($\AA$)$&$\gamma_{\phi}(^\mathrm{\circ})$&$\sigma_\phi(^\mathrm{\circ})$&$\xi_{\phi}($\AA$)$&$\gamma_{\omega}(^\mathrm{\circ})$&$\sigma_\omega(^\mathrm{\circ})$&$\xi_{\omega}($\AA$)$ \\ 
\hline
Inner Peak & 49.09 & 0.004(3) & 0.0035 & 228 & 1.(1) & 0.2340 & 425 & 0.2(3) & 0.1511 & 1923\\  
\hline
Outer Peak & 25.75 & 0.01(2) & 0.0067 & 80 &  0.8(8) &  0.178 & 265 & 0.1(0) & 0.287 & 2173\\
\hline \hline
\end{tabular}
\end{table*}

\section{Relationship between peaks at $Q_{1}$ and previously observed peaks in Phase IV}\label{schematic_ps}

\renewcommand{\thefigure}{D\arabic{figure}}
\renewcommand{\thetable}{D\Roman{table}}
\setcounter{figure}{0}
\setcounter{table}{0}

\begin{figure}[h]
\includegraphics[width= 0.5\columnwidth] {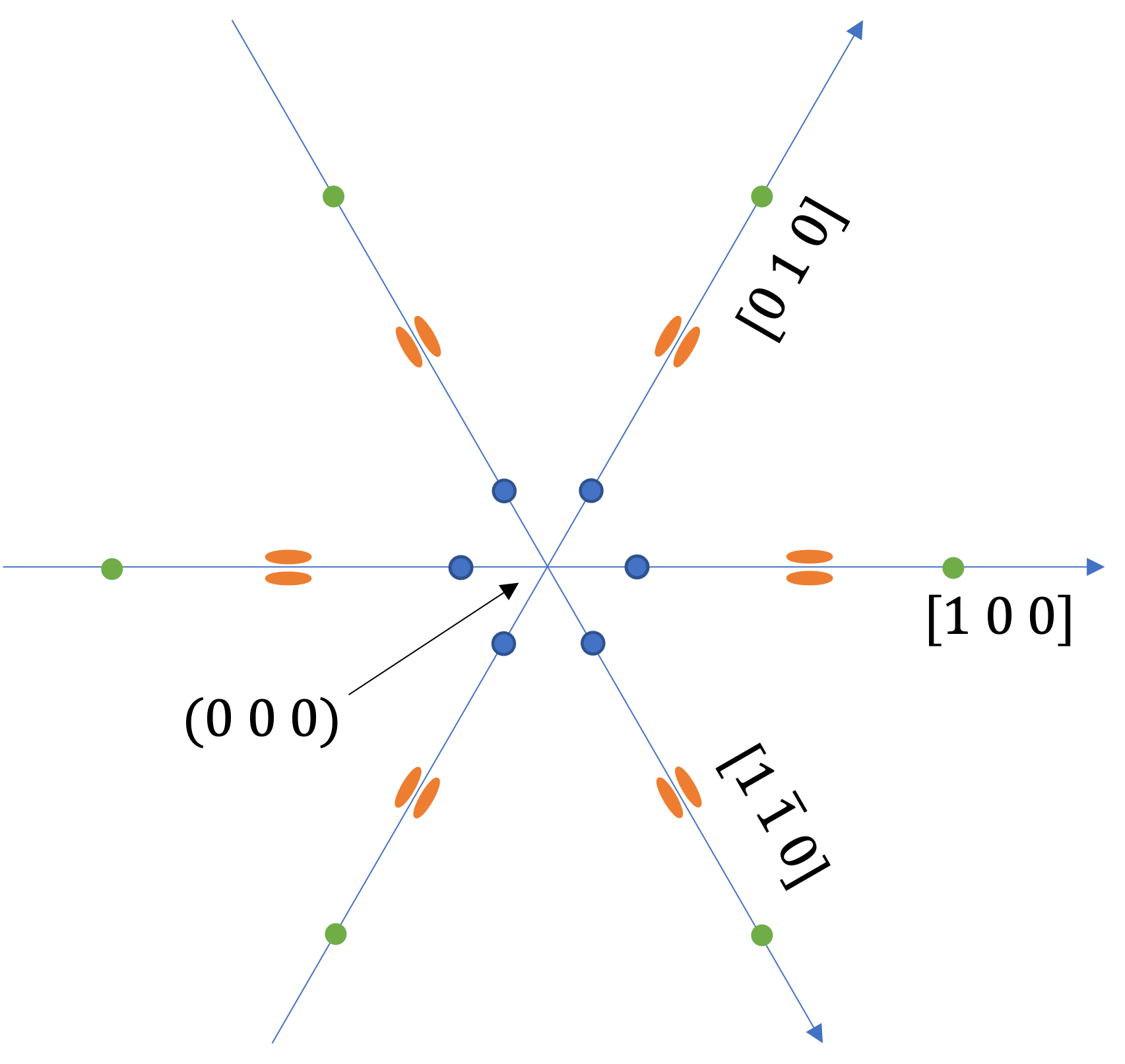}
\caption{\label{schematic_peaks}Schematic diagram showing the relationship between the newly observed peaks at $Q_{1}$ (blue) and the previously reported peaks at higher $Q$s as studied by conventional neutron diffraction in higher Brillouin zones\cite{Lebech_1994} in phase IV.  As reported in the literature, the orange tangentially split ovals at $Q$ $\approx$ 0.21\AA $^{-1}$ correspond to magnetic ordering  of the hexagonal Nd sites while the modulation vectors represented by the green circles ($Q$ $\approx$ 0.36\AA$^{-1}$) correspond to ordering of the cubic Nd sites \cite{Lebech_1994}. The relative appearance of the $Q_{1}$ and the peaks from the hexagonal ordering in the SANS data is shown in Fig. \ref{pretty_with_ferro}(a).  Note the peaks correspond the ordering of the cubic sites are out of the region of reciprocal space probed in the SANS measurements reported here.}
\end{figure}

\section{Magnetometry and thermal hysteresis}\label{mag_thermal}

\renewcommand{\thefigure}{E\arabic{figure}}
\renewcommand{\thetable}{E\Roman{table}}
\setcounter{figure}{0}
\setcounter{table}{0}

\begin{figure}[h]
\includegraphics[width= 0.5\columnwidth] {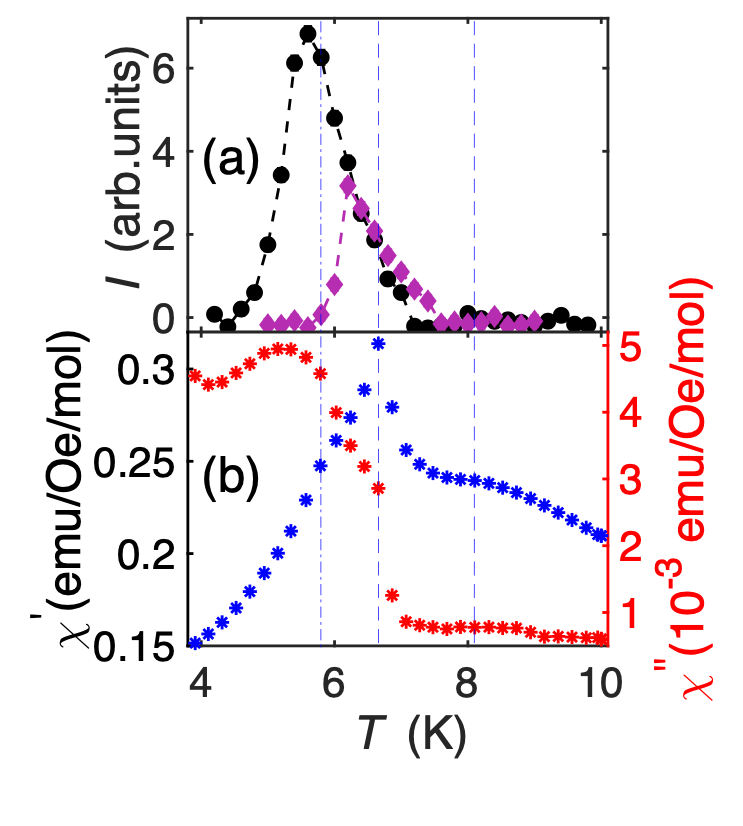}
\caption{
Panel (a) shows the intensity at $Q_{1}$ as a function of temperature for cooling (circles) and warming (diamonds) under zero applied field.   Panel (b) shows the in-phase, $\chi'$, (blue) and out of phase, $\chi''$ (red) A.C. susceptibility data collected on cooling with an applied field of $H_{AC} = 2~ Oe$, frequency, f = 21 Hz, and $H_{DC} = 0 ~Oe$. The vertical dotted lines indicate anomalies in the first derivative of the in-phase component of the A.C. susceptibility at T = 5.7(9) K, 6.6(6) K,and 8.1(0) K. Data was collected for a single crystal that was cut from a larger crystal used for SANS, polished to clean the surface of oxide, wrapped in Ta foil, and then annealed for 12h at 700C in a vacuum-sealed silica ampoule.  
\label{hysteresis_neu} }
\end{figure}

\begin{figure}[h]
\includegraphics[width= 0.4\columnwidth] {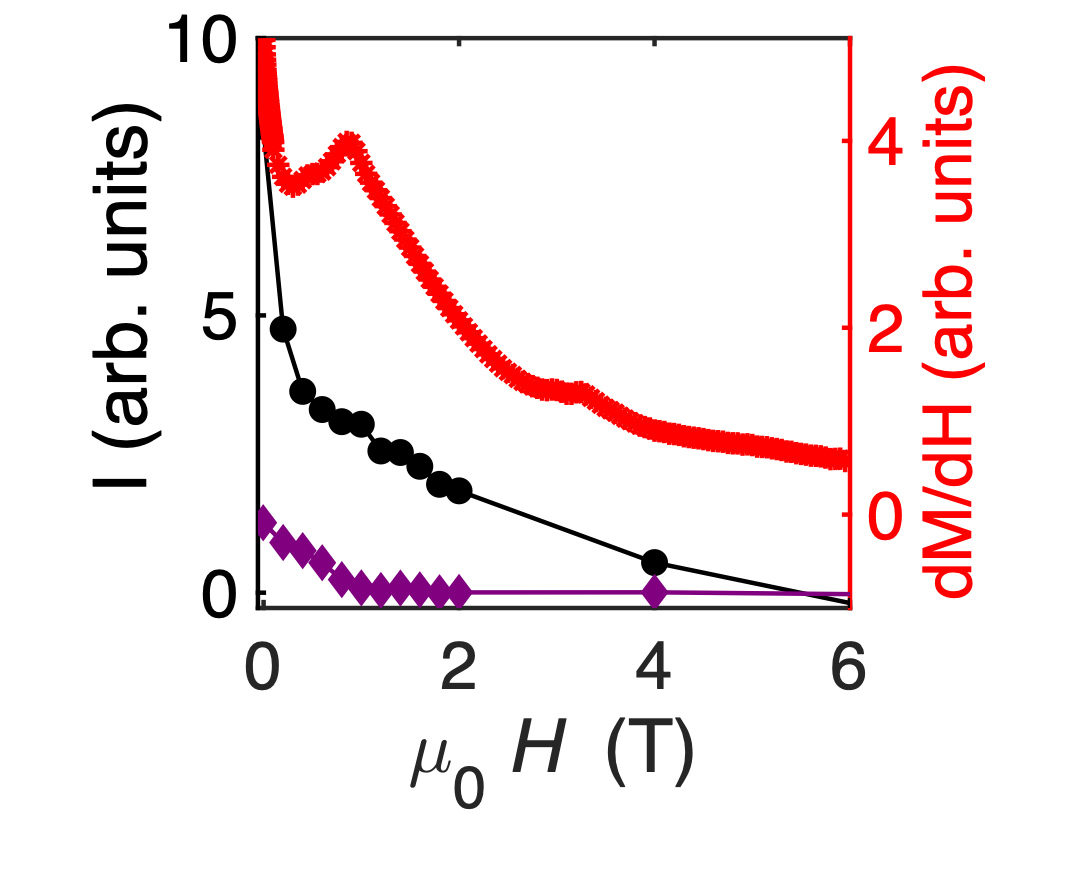}
\caption{\label{hex_fs}
The field dependence of the integrated intensity of the $Q_{1}$ peaks (purple) and the low-$Q$ scattering integrated for 0.001 \AA$^{-1} \leq Q \leq 0.056$ \AA$^{-1}$ (black) at 6.5 K (Phase IV). The red curve shows the derivative of isothermal magnetization $(dM/dH)$ measurements performed at T = 6.6 K with increasing field applied along the c-axis. The $(dM/dH)$ data have an anomaly at 0.85 T that approximately coincides with the phase boundary determined from the $Q_{1}$ peaks at 1.1 T. Data was collected on the same sample as in Fig. \ref{hysteresis_neu}.  
}
\end{figure}

\begin{figure}[h]
\includegraphics[width= 0.4\columnwidth] {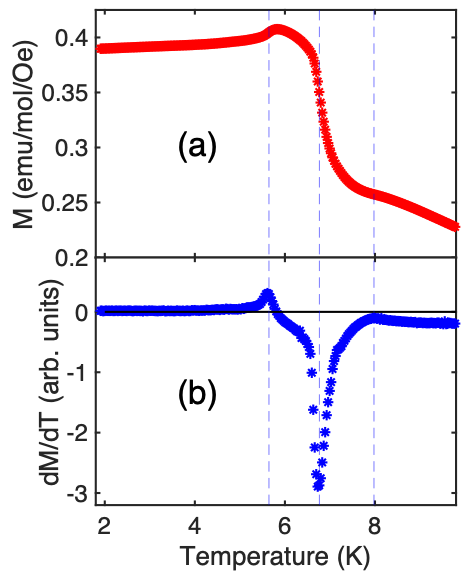}
\caption{\label{dc_sus}
 Panel (a) shows the temperature dependence of the magnetization while field cooling with a field of 10 Oe. A noticeable enhancement in the magnetization is observed at temperatures below  $\approx$ 8 K with anomalies in $dM/dT$ occurring at 7.9(8), 6.7(6) K and 5.6(5) K. (b) $dM/dT$ emphasises  the aforementioned changes in the magnetic susceptibility. Data was collected on the same sample as in Fig. \ref{hysteresis_neu}.
}
\end{figure}

Figure \ref{hysteresis_neu}(a) shows the integrated intensity of the peaks at $Q_{1}$ on warming and cooling.  The significant difference between the curves provides a clear indication of the first order transition into phase IV as previously reported\cite{Lebech_1994}.  Additionally, as indicated by the dashed lines in Fig. \ref{hysteresis_neu}(b), the first derivative of the in-phase component of the ac susceptibility exhibits extrema at 5.7(9) K, and 8.1(0) K.  At  6.6(6) K, there is a zero crossing of the first derivative as the gradient on the two sides are opposite in sign. As described in the main text, the scattering integrated between 0.001 \AA$^{-1}$ and  0.056 \AA$^{-1}$ is strong in the region where phase IV is observed.  To further explore the association of this scattering with Phase IV, bulk magnetometry data is shown in Figs. \ref{hex_fs} and \ref{dc_sus}.  The data in Fig. \ref{hex_fs} indicates an anomaly near $\mu_0H$ = 0.9 T which is near the phase boundary determined from the neutron scattering data and the previously reported magnetization data of \cite{boghosian_1985} which exhibits an inflection point near 1 T.

 The field dependence of the $Q_{1}$ modulations is shown in fig. \ref{field_dep}(d). These data indicate a phase boundary at $\mu_0H$ = 1.(1) T. Moreover, Fig. \ref{hex_fs} shows the field dependence of the bulk susceptibility and the low $Q$ scattering originating from probable ferromagnetic correlations present in phase IV, alongside the field dependence of the $Q_{1}$ modulations. Notably in the region where the $Q_{1}$ scattering is present, the ferromagnetic scattering is enhanced and the bulk magnetic susceptibility also indicates an anomaly at 0.9 T.
 
 \newpage{}

\newpage{}

\section{Synchrotron x-ray measurements}\label{x_raya}

\renewcommand{\thefigure}{F\arabic{figure}}
\renewcommand{\thetable}{F\Roman{table}}
\setcounter{figure}{0}
\setcounter{table}{0}

Synchrotron x-ray diffraction measurements in a magnetic field were performed at the 6-ID-C experimental station of the Advanced Photon Source (APS). A single-crystal Nd sample was mounted at the tip of a He-4 flow cryostat (Variable Temperature Insert, VTI) for measurements at low temperatures in such a way that its face was parallel to the VTI axis to enable reflection geometry. The VTI was in turn inserted into a split-pair, vertical-field superconducting magnet with a maximum field of 4.5 T. The magnet was mounted on a two-circle horizontal diffractometer with a limited chi circle motion of 3.4$^{\circ}$. In this geometry, the magnetic field was parallel to sample surface. A Si (111) double-bounce monochromator is used to select a photon energy with a 0.01 \% bandwidth. For the diffraction experiment, we used 20 keV x-rays. The second crystal of monochromator was detuned to suppress higher harmonic contamination. The horizontal size of the incident beam is controlled by slits in front of the sample. A point detector (NaI-scintillator) was used for recording the scattering intensity of charge Bragg peaks. The incident beam intensity was measured with a N$_2$-gas filled ion chamber.

\begin{figure}[h]
\includegraphics[width= .95\columnwidth] {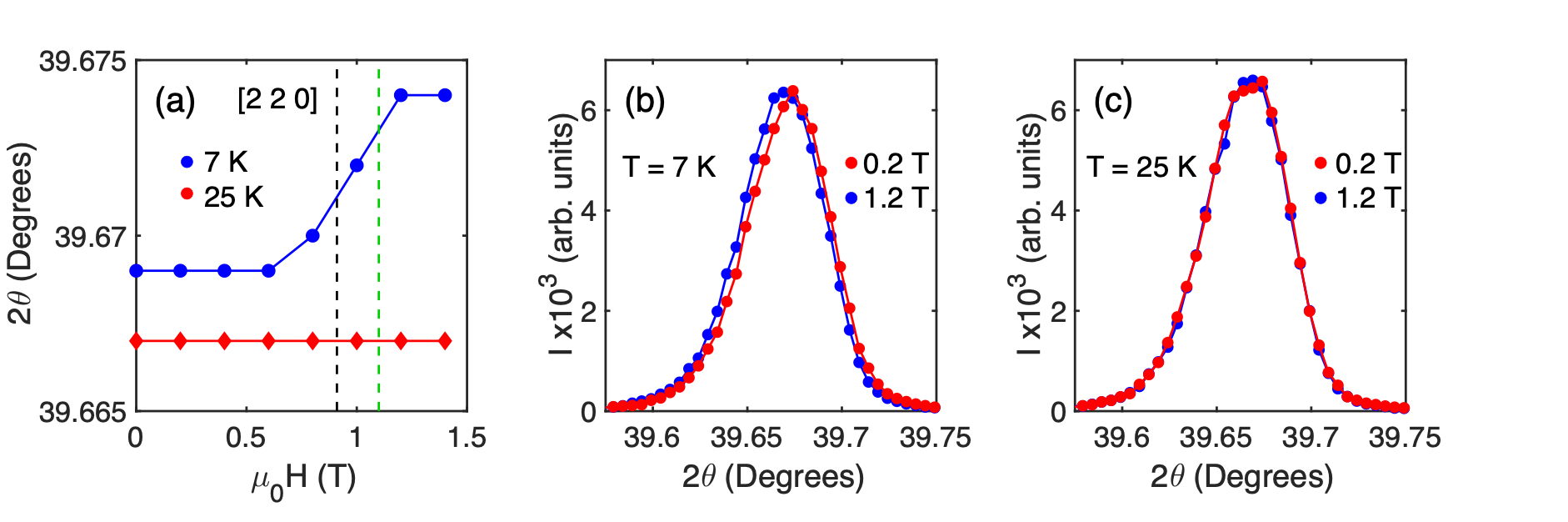}
\caption{\label{xray} Synchrotron x-ray diffraction data showing the response of the lattice to applied magnetic fields.   (a)  Scattering angle (2$\theta$) dependence of the (2 2 0) structural Bragg peak at 7 and 25 K. The phase boundary identified by the SANS measurements near 1 T is indicated by the dashed line.  (b) Scans of the (2 2 0) structural Bragg peak at 0.2 and 1.2 T at 7 K.  (c)    Scans of the (2 2 0) structural Bragg peak at 0.2 and 1.2 T at 25 K.
}
\end{figure}

\newpage{}

\section{Hall effect measurements}

\renewcommand{\thefigure}{G\arabic{figure}}
\renewcommand{\thetable}{G\Roman{table}}
\setcounter{figure}{0}
\setcounter{table}{0}

\begin{figure}[h]
\includegraphics[width= 0.45\columnwidth] {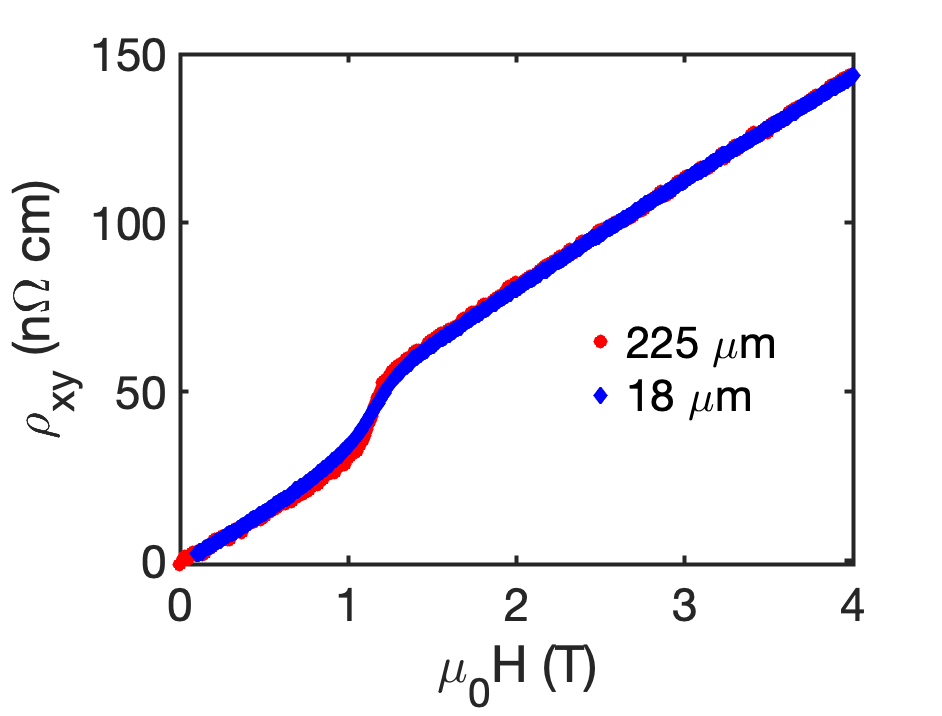}
\caption{\label{Hall_diff_thickness} Hall effect measurements at 2 K.  The plot shows there is a broadened anomaly in the Hall signal in both samples corresponding to the phase transition near 1 T (see Fig. \ref{struct}).  The samples have thicknesses of 225 $\mu$m (red) and 18 $\mu$m (blue). The transition is sharper in the thicker sample. For the Hall measurements, the current was applied along the a-axis; the voltage was measured along the a* direction; and the magnetic field was applied along the c-axis. Currents of 10 mA and 3.6 mA were used for the 225 $\mu$m and 18 $\mu$m samples respectively.}
\end{figure}

 The Hall effect was measured in mechanically thinned samples of Nd. Samples were mechanically thinned to 15~$\mu$m 225~$\mu$m. In the 15~$\mu$m sample, we observed that the magnetic transition near 1 T at 2 K (Phase VI) was extremely broadened and occurred over a field range of 0.8 T (Fig. \ref{Hall_diff_thickness}). We then measured a mechanically thinned 225~$\mu$m sample instead, which also showed a broadened transition occurring over a field range of 0.4 T (Fig. \ref{Hall_diff_thickness}). No evidence for a topological hall effect was found in either sample.  The lack of topological Hall effect may then imply that there is no topologically no trivial spin texture present, the sample properties are sufficiently changed by the mechanical thinning process, or possibly that a more complex spin texture is present that wouldn't exhibit a topological hall effect\cite{afskyrmion_2017}.  

\end{document}